\documentclass[final,a4paper,USenglish]{lipics}

\usepackage{microtype}

\usepackage{amssymb}
\usepackage{amsmath}
\usepackage{bussproofs}
\usepackage{caption}
\usepackage[noadjust]{cite}
\usepackage{eufrak}
\usepackage{graphicx}
\usepackage{mdwlist}
\usepackage{subfig}
\usepackage{ulem}
\usepackage{url}
\usepackage{tikz}
\usetikzlibrary{positioning,fit,arrows,decorations.markings}
\usepackage{xspace}

\newcommand{\jambox}{\ensuremath{\mathsf{Jambox}}\xspace}

\newcommand{\DP}{\mathit{DP}}

\DeclareSymbolFont{letters}{OML}{cmbboard}{m}{it} 

\newcommand{\from}{\leftarrow}
\def\test#1#2#3{\setbox0=\hbox{$\vphantom{#1}^{#2}_{#3}$}%
                \dimen0=\wd0%
                \setbox1=\hbox{$\scriptstyle #2$}%
                \advance\dimen0-\wd1%
                \setbox1=\hbox{\hskip\dimen0\copy1}%
                \dimen0=\wd0%
                \setbox2=\hbox{$\scriptstyle #3$}%
                \advance\dimen0-\wd2%
                \setbox2=\hbox{\hskip\dimen0\copy2}%
                {\vphantom{#1}^{\box1}_{\box2}}{#1}
}

\newcommand\RR{\ensuremath{\mathcal{R}}\xspace}

\newcommand\PP{\ensuremath{\mathcal{P}}\xspace}

\newcommand{\rEX}[1]{Example~\ref{#1}}

\newcommand{\rLE}[1]{Lemma~\ref{#1}}

\newcommand\relR{\mathcal{R}}
\newcommand\relS{\mathcal{S}}
\newcommand\dpP{\mathcal{P}}
\newcommand\dpR{\mathcal{R}}
\newcommand\rdpP{\mathcal{P}}
\newcommand\rdpR{\mathcal{R}}
\newcommand\rdpPw{\mathcal{P}_w}
\newcommand\rdpRw{\mathcal{R}_w}
\newcommand\dpp{(\dpP,\dpR)}
\newcommand\rdpp{(\rdpP,\rdpPw,\rdpR,\rdpRw)}
\newcommand\rtp{\relR / \relS}
\newcommand\aux{\mathit{aux}}

\newcommand{\Fa}{\mathsf{a}}

\newcommand{\Fb}{\mathsf{b}}

\newcommand{\sep}{\hspace{-0.28em}}

\newlength{\rtimesl}
\newcommand{\overlay}%
{\mathrel{\from\sep{\ltimes\hspace{-\rtimesl}\rtimes}\sep\to}}

\newcommand\QQ{\mathcal{Q}}

\newcommand\TTTT{%
 \ensuremath{\mathsf{T\kern-0.2em\raisebox{-0.3em}%
 {$\mathsf{T}$}\kern-0.2emT\kern-0.2em%
 \raisebox{-0.3em}$\mathsf{2}$}}\xspace%
}

\newcommand\isafor{\textsf{Isa\kern-0.2exF\kern-0.2exo\kern-0.2exR}\xspace}
\newcommand\ceta{\textsf{C\kern-0.2exe\kern-0.5exT\kern-0.5exA}\xspace}

\newcommand{\FFF}{\mathsf{F}}

\newcommand\sn{\textsf{SN}}

\newcommand\stackarr[2][]{\stackrel{%
  \makebox[0pt]{\raisebox{0pt}[0pt][0pt]{\(\scriptstyle#1\)}}}{\to}_{#2}}
\newcommand\rstep[1][XXX]{\stackarr{#1}^{}}

\newcommand\rsteps[1][\RR]{\stackarr{#1}^*}

\title{A Relative Dependency Pair Framework%
\footnote{%
  This research is supported by the Austrian Science Fund (FWF): P22767, J3202.}}

\author[1]{Christian Sternagel}
\author[2]{Ren{\'e} Thiemann}

\affil[1]{School of Information Science\\
  Japan Advanced Institute of Science and Technology, Japan\\
  \texttt{c-sterna@jaist.ac.jp}}
\affil[2]{Institute of Computer Science\\
  University of Innsbruck, Austria\\
  \texttt{rene.thiemann@uibk.ac.at}}

\Copyright[nc-nd]
          {C.~Sternagel, R.~Thiemann}
\theoremstyle{plain}
\theoremstyle{definition}

\newcommand{\toe}{\stackrel\epsilon\to}

\newcommand\comp{\cdot}

\newcounter{rewriterule}
\begin{document}
\maketitle


\section{Introduction}

Relative rewriting and the dependency pair framework (DP framework) are two strongly 
related termination methods. In both formalisms we consider 
whether the combination of two TRSs allows an infinite derivation:
\begin{itemize} 
\item Relative termination of $\rtp$ can be defined as strong normalization
 of ${\to_{\relR}} \comp {\to_{\relS}^*}$.
\item Finiteness of a DP problem $\dpp$ can be defined as strong normalization
 of ${\toe_{\dpP}} \comp {\to_{\dpR}^*}$ where $\toe$ allows steps only at
 the top.
 Moreover, \emph{minimality} can be incorporated by requiring
 that all terms are terminating w.r.t.\ $\dpR$.
\end{itemize}

The above definitions have two orthogonal distinctions of rules.
In both formalisms there are \emph{strict} and \emph{weak} rules: 
$\dpP$ and $\relR$ are the strict rules of $\dpp$ and $\rtp$, respectively,
while
$\dpR$ and $\relS$ are the respective weak rules.
In the DP framework, there is the additional distinction between rules that may only be applied at the 
top ($\dpP$) and those that can be applied at arbitrary positions ($\dpR$).

Note that the restriction to top rewriting is an important advantage for proving termination
in the DP framework.
It allows to use non-monotone orders for orienting the strict rules.
Furthermore, if minimality is considered, we can use termination techniques
(e.g., usable rules or the subterm criterion) that are not available for relative rewriting.

\def\fstRtp{(\relR_s \cup \relS_s) / (\relR_w \cup \relS_w)}
\def\sndRtp{\relR_w / \relS_w}
However, also relative rewriting has some advantages which are currently not
available in the DP framework: Geser showed that it is always possible to split
a relative termination problem into two parts \cite{Geser90}.
Relative termination of $(\relR_s \cup \relR_w) / (\relS_s \cup \relS_w)$ can be
shown by relative termination of both $\fstRtp$ and $\sndRtp$.  Hence, it is
possible to show (in a first relative termination proof) that the strict rules
$\relR_s \cup \relS_s$ cannot occur infinitely often and afterwards continue (in
a second relative termination proof) with the problem $\sndRtp$.  A major
advantage of this approach is that in the first proof we can apply arbitrary
techniques which may increase the size of the TRSs drastically (e.g., semantic
labeling \cite{Z95}), or which may even be incomplete (e.g., string reversal in
combination with innermost rewriting, where by reversing the rules we have to
forget about the strategy). As long as relative termination of $\fstRtp$ could
be proven, we can afterwards continue independently with the problem $\sndRtp$.

Such a split is currently not possible in the DP framework since there are no
top weak rules and also no strict rules which can be applied everywhere.

In this paper we generalize the DP framework to a relative DP framework, where such
a split is possible. To this end, we consider DP problems of the form
$\rdpp$, where
we have strict and weak, top and non-top rules. (This kind of DP problems were
first suggested by J\"org Endrullis at the \emph{Workshop on the Certification of
Termination Proofs} in 2007 and are already used in his termination tool
\jambox \cite{jambox}. Unfortunately the suggestion did not get much
attention back then and we are not aware of any publications on this topic.)
In this way, problems that occur in combination with semantic
labeling and dependency pairs---which can otherwise be solved by using a dedicated semantics for DP problems \cite{SemlabFW}---can
easily be avoided. Furthermore, the new framework is more general than
\cite{SemlabFW} since it also solves some problems
that occur when using other termination techniques like uncurrying \cite{UncurryHMZ,UncurryST}.

\section{A Relative Dependency Pair Framework}

We assume familiarity with term rewriting \cite{baader-nipkow} and the DP framework \cite{LPAR04}.
\begin{definition}
A \emph{relative dependency pair problem} $\rdpp$ is a quadruple of TRSs
with \emph{pairs} $\rdpP \cup \rdpPw$ (where pairs from $\rdpP$ are called
\emph{strict} and those of $\rdpPw$ \emph{weak}) and \emph{rules} $\rdpR \cup
\rdpRw$ (where rules of $\rdpR$ are called \emph{strict} and those of $\rdpRw$
\emph{weak}).
\end{definition}
For relative DPPs the notion of chains and finiteness is adapted in the
following way.
\begin{definition}
\label{reldp}
An infinite sequence of pairs
$s_1 \to t_1$, $s_2 \to t_2$, \ldots\ forms a \emph{$\rdpp$-chain} if there
exists a corresponding sequence of substitutions $\sigma_1$, $\sigma_2$,
\ldots\ such that
\begin{gather}
s_i \to t_i \in \rdpP \cup \rdpP_w\ \text{for all $i$}\\
t_i\sigma_i \rsteps[\rdpR\cup\rdpRw] s_{i+1}\sigma_{i+1}\ \text{for all $i$}\\
s_i \to t_i \in \rdpP\ \text{or}\ 
t_i\sigma_i \rsteps[\rdpR\cup\rdpRw] \comp \rstep[\rdpR] \comp \rsteps[\rdpR\cup\rdpRw]
s_{i+1}\sigma_{i+1}\ \text{for infinitely many $i$}
\intertext{For minimal chains, we additionally require}
\sn_{\rdpR\cup\rdpRw}(t_i\sigma_i)\ \text{for all $i$}
\end{gather}
A
relative DPP $\rdpp$ is \emph{finite}, iff there is
no minimal infinite $\rdpp$-chain.
\end{definition}

Hence, a (minimal) $\rdpp$-chain is like a (minimal) $(\rdpP \cup \rdpPw,\rdpR
\cup \rdpRw)$-chain%
---as defined in \cite{AG00}---with 
the additional demand that there are infinitely many strict steps using $\rdpP$
or $\rdpR$.
It is easy to see that $\dpp$-chains can be expressed in the new framework. 

\begin{lemma}
The DP problem $\dpp$ is finite iff there exists a minimal $\dpp$-chain iff
there exists a minimal $(\dpP,\emptyset,\emptyset,\dpR)$-chain iff the relative DPP
$(\dpP,\emptyset,\emptyset,\dpR)$ is finite.
\end{lemma}

Note that in contrast to DPPs $\dpp$, for relative DPPs, $\rdpP = \emptyset$
does not imply finiteness of $\rdpp$.
\begin{example}
The relative DPP $(\emptyset,\{\FFF(\Fa) \to \FFF(\Fb) \}, \{\Fb \to \Fa\},
\emptyset)$ is not finite.
\end{example}
However, a sufficient criterion for finiteness is that 
there are either no pairs, or that there are neither strict pairs nor strict rules.

\begin{lemma}[Trivially finite relative DPPs]
\label{trivial}
If $\rdpP \cup \rdpPw = \emptyset$ or $\rdpP \cup \rdpR = \emptyset$ then $\rdpp$ is finite.
\end{lemma}

\section{Processors in the Relative Dependency Pair Framework}

Processors and soundness of processors in the relative DP framework are defined
as in the DP framework, but operate on relative DPPs instead of
DPPs (a processor is sound if finiteness of all resulting relative DPPs implies
finiteness of the given relative DPP).

Note that most processors can easily be adapted to the new framework where most often it suffices
to treat the relative DPP $\rdpp$ as the DPP $(\rdpP \cup \rdpPw,\rdpR \cup
\rdpRw)$.

However, when starting with the initial relative DPP $(\DP(\RR),\emptyset,\emptyset,\RR)$ it is questionable
whether we ever reach relative DPPs containing weak pairs or strict rules. If this is not the case,
then our generalization would be useless. Therefore, in the following we give evidence that the additional flexibility 
is beneficial. 

Easy examples are semantic labeling and uncurrying. Both techniques are transformational techniques 
where each original 
step is transformed into one main transformed step together with some auxiliary steps. For the
auxiliary steps one uses auxiliary pairs and rules (the decreasing rules and the
uncurrying rules, respectively).
If there are auxiliary pairs $\PP_{\aux}$, then in the DP framework, $\PP_{\aux}$ can only be added as strict pairs,
whereas in the relative DP framework, we can add $\PP_{\aux}$ to the weak pairs,
and hence we do
not have to delete all pairs of $\PP_{\aux}$ anymore for proving finiteness.

As another example, we consider top-uncurrying of \cite[Def.~19]{UncurryST}, where some rules $\RR$ are turned into pairs.
Again, in the DP framework this would turn the weak rules $\RR$ into strict pairs, which in fact would
demand that we prove termination of $\RR$ twice: Once via the original DPs for $\RR$, and a second time
after the weak rules of $\RR$ have been converted into strict pairs. For example, in \cite[Ex.~21]{UncurryST} termination
of the minus-rules is proven twice. This is no longer required in the relative
DP framework where one can just turn the weak rules $\RR$ into weak pairs $\RR$.

Finally, in the relative DP framework we can apply the split technique known
from relative rewriting.

\begin{definition}[Split processor]
The relative DPP $(\PP^1_s \cup \PP^1_w,\PP^2_s \cup \PP^2_w,\RR^1_s \cup \RR^1_w,\RR^2_s \cup \RR^2_w)$ is finite 
if both $(\PP^1_s \cup \PP^2_s, \PP^1_w \cup \PP^2_w, \RR^1_s \cup \RR^2_s, \RR^1_w \cup \RR^2_w)$ and 
$(\PP^1_w,\PP^2_w,\RR^2_w,\RR^2_w)$ are finite.
\end{definition}

A more instructive way of putting the above definition for termination tool
authors that are used to standard DP problems is as follows.  Start from the
relative DPP $(\PP,\emptyset,\emptyset,\RR)$. Identify pairs $\PP'$ and rules
$\RR'$ that should be deleted. Then use the split processor to obtain the two
relative DPPs $(\PP',\PP\setminus\PP',\RR',\RR\setminus\RR')$ and
$(\PP\setminus\PP',\emptyset,\emptyset,\RR\setminus\RR')$.

Clearly, the split processor can be used to obtain relative DPPs with strict rules and weak pairs, but the question
is how to apply it. We give two possibilities.

\subparagraph{Semantic labeling} is often used in a way, that after labeling one tries to remove all labeled variants of some
  rules $\RR_s$ and pairs $\PP_s$, and afterwards removes the labels again to
  continue on a smaller unlabeled problem. 
  
\begin{example}
\label{exsem}
Consider a DP problem $p_1 = (\{1,2\},\{3\})$.
After applying semantic labeling, all pairs and rules occur in labeled 
variants $1.x$, $2.x$, and $3.x$, so the resulting DP problem might look
like $(\{1.a,1.b,2.a,2.b\},\{3.a,3.b,3.c\})$. Applying standard techniques to remove
pairs and rules one might get stuck at $p_2 = (\{2.a,2.b\},\{3.a,3.c\})$. Although
$p_1$ contains less rules than $p_2$, $p_2$ is somehow simpler since all rules $1.x$
have been removed. And indeed, after applying unlabeling on $p_2$ the resulting
DP problem $p_3 = (\{2\},\{3\})$ is smaller than $p_1$.
\end{example}

  Since the removal of labels is problematic for soundness,
  a special semantics was developed in \cite{SemlabFW}. This is no longer required in the relative DP framework.
  After $\RR_s$ and $\PP_s$ have been identified, one just applies the split processor to transform
  $(\PP,\emptyset,\emptyset,\RR)$ into $(\PP_s,\PP \setminus \PP_s,\RR_s, \RR \setminus \RR_s)$ and 
  $(\PP \setminus \PP_s, \emptyset, \emptyset, \RR \setminus \RR_s)$. The proof that all labeled variants of rules
  in $\RR_s$ and pairs in $\PP_s$ can be dropped, proves finiteness of the first problem, and one can continue on the latter
  problem without having to apply unlabeling.
  
  \begin{example}
Using split, we can restructure the proof of \rEX{exsem} without using unlabeling:
We know that in the end, we only get rid of pair $1$. Hence, we apply split
on $p_1$ to obtain $p_3$ and $p_4 = (\{1\},\{2\},\emptyset,\{3\})$. Thus,
we get the same remaining problem $p_3$ if we can prove finiteness of $p_4$.
But this can be done by replaying the proof steps in \rEX{exsem}.
Applying the same labeling as before, we obtain $p_5 = (\{1.a,1.b\},\{2.a,2.b\},
\emptyset,\{3.a,3.b,3.c\})$. Removing pairs and rules as before, we simplify
$p_5$ to $p_6 = (\emptyset,\{2.a,2.b\},\emptyset,\{3.a,3.c\})$ and this relative
DP problem is trivially finite by \rLE{trivial}.
\end{example}
  
  Note that using \cite{SemlabFW} it was only possible to revert the labeling, but not to revert
  other techniques like the closure under flat contexts which is used in
  combination with root-labeling \cite{root-labeling}. 
  However, using the split processor this is also easily possible, since one just has to apply the split processor
  before applying the closure under flat contexts.
  
  A further advantage of the relative DP framework in comparison to \cite{SemlabFW}
  can be seen in the combination of semantic labeling with the dependency graph 
  processor.
  
\begin{example}
 Consider a DP problem $p_1 = (\{1,2\},\{3,4\})$ which is
 transformed into $(\{1.a,1.b,2.a,2.b\},\{3.a,3.b,4.a,4.b\})$ using semantic
 labeling.
 Applying the dependency graph and reduction pairs yields two remaining
 DP problems $p_2 = (\{2.a\},\{4.a\})$ and $p_3 = (\{2.b\},\{3.a,4.b\})$.
 Using unlabeling we have to prove finiteness of the two
 remaining problems $p_4 = (\{2\},\{4\})$ and $p_5 = (\{2\},\{3,4\})$.
 Note that finiteness of $p_5$ does not imply finiteness of $p_4$, so 
 one indeed has to perform two proofs.
 
 However, when using the split processor, only $p_5$ remains: we observe from
 $p_2$ and $p_3$ that
 only pair $1$ could be removed. So, we start to split $p_1$ into
 $p_5$ and $p_6 = (\{1\},\{2\},\emptyset,\{3,4\})$.
 Labeling $p_6$ yields $(\{1.a,1.b\},\{2.a,2.b\},\emptyset,\{3.a,3.b,4.a,4.b\})$
 which is simplified to the two problems 
 $(\emptyset,\{2.a\},\emptyset,\{4.a\})$
 and 
 $(\emptyset,\{2.b\},\emptyset,\{3.a,4.b\})$ with the same techniques as before.
 Both problems are trivially finite by \rLE{trivial}.
\end{example}

\subparagraph{Other Techniques} may also take advantage of the split processor. For example, the dependency pair
  transformation of narrowing \cite{AG00,LPAR04} is not complete in the innermost case but might help to 
  remove some pairs and rules. If it turns out that after some narrowing steps some original pairs and rules can
  be removed, then one can just insert a split processor before narrowing has been performed. In this way one has 
  obtained progress in proving finiteness and in the remaining system the potential incomplete narrowing steps 
  have not been applied. In other words, the split processor allows to apply incomplete
  techniques without losing overall completeness.

\section{Conclusions and Future Work}

We presented the relative DP framework which generalizes the existing DP framework
by allowing weak pairs and strict rules.
It forms the basis of our proof checker \ceta (since version 2.0) \cite{ceta-tphols}
where we additionally 
integrated innermost 
rewriting (in the form of $\QQ$-restricted rewriting) \cite{LPAR04}. 
One of the main features of the new framework is the possibility to split
a DP problem
into two DP problems which can be treated independently. 
Examples to illustrate the new features are provided in the \isafor-repository
(e.g., \href{http://cl-informatik.uibk.ac.at/software/ceta/examples/div_uncurry.proof.xml}{\texttt{div\_uncurry.proof.xml}} uses weak pairs
for uncurrying,
and in \href{http://cl-informatik.uibk.ac.at/software/ceta/examples/secret_07_trs_4_top.proof.xml}{\texttt{secret\_07\_trs\_4\_top.proof.xml}}
the split processor is used to avoid unlabeling). 

It is an obvious question, whether the relative DP framework can
be used to characterize relative termination. In a preliminary version we 
answered this question positively by 
presenting a theorem that $\relR / \relS$ is relative terminating iff there is 
no infinite $(\DP(\relR),\DP(\relS),\relR,\relS)$-chain. 
However, it was detected that the corresponding proof contained
a gap (it was the only proof that we did not formalize in Isabelle/HOL) and
that the whole theorem did not hold (by means of a counterexample).

An interesting direction for future work is to unify termination (via relative
DP problems) with relative termination. One reason is that this would allow to
reduce the formalization effort, since results for termination are expected to
be corollaries carrying over from relative termination.

\subparagraph{Acknowledgments}
We would like to thank J\"org Endrullis and an anonymous referee for pointing out
that our attempt to characterize relative termination using
the relative DP framework is unsound.
It remains as interesting open problem to give such a characterization.

\bibliographystyle{plain}
\bibliography{references}

\end{document}